\documentclass{article} 
\usepackage{iclr_aims,times}

\usepackage{amsmath,amsfonts,bm}

\def\eqref#1{equation~\ref{#1}}

\def\1{\bm{1}}

\DeclareMathAlphabet{\mathsfit}{\encodingdefault}{\sfdefault}{m}{sl}
\SetMathAlphabet{\mathsfit}{bold}{\encodingdefault}{\sfdefault}{bx}{n}

\usepackage{hyperref}
\usepackage{url}
\usepackage{multirow}
\usepackage{cleveref}
\usepackage{graphicx}
\usepackage{pifont}
\usepackage{tabularx}
\usepackage{makecell}
\usepackage{booktabs, multirow} 
\usepackage{todonotes}
\usepackage{setspace}
\usepackage{glossaries}

\newacronym{fwp}{FWP}{Few-Shot Prompting}
\newacronym{pe}{PE}{Prompt Engeneering}
\newacronym{fl}{FL}{Federated Learning}
\newacronym{noniid}{Non-IID}{Non-Independently Identically Distributed}
\newacronym{cd}{CD}{Concept Drift}
\newacronym{fedavg}{FedAvg}{Federated Average}
\newacronym{llm}{LLM}{Large Language Model}
\newacronym{poc}{PoC}{Power-of-Choice}
\newacronym{pooc}{ProC}{Proof-of-Concept}
\newacronym{ai}{AI}{Artificial Intelligence}
\newacronym{ml}{ML}{Machine Learning}
\newacronym{mas}{MAS}{Multi-agent System}
\newacronym{lmagents}{LM-Agents}{Language Model Agents}
\newacronym{rag}{RAG}{Retrieval-Augmented Generation}
\newacronym{lm}{LM}{Language Model}
\newacronym{pfl}{PFL}{Personalized Federated Learning}
\newacronym{dp}{DP}{Differential Privacy}
\newacronym{he}{He}{Homomorphic Encryption}
\newacronym{ra}{RA}{Robust Aggregation}
\newacronym{cot}{CoT}{Chain-of-Thought}
\newacronym{tot}{ToT}{Tree-of-Thought}
\newacronym{pot}{PoT}{Proof-of-Thought}
\newacronym{got}{GoT}{Graph-of-Thought}
\newacronym{sml}{SML}{Small Language Model}
\newacronym{iot}{iot}{Internet of Things}

\title{Agentic Federated Learning: The Future of Distributed Training Orchestration}
\author{Rafael O. Jarczewski\thanks{Institute of Computing, University of Campinas and Hub of Artificial Intelligence and Cognitive Architectures - H.IAAC, Campinas, Brazil} \And Gabriel U. Talasso$^{\star}$ \And Leandro Villas$^{\star}$ \And Allan {M. de Souza}$^{\star}$
}

\iclrfinalcopy 

\begin{document}

\maketitle

\begin{abstract}
Although Federated Learning (FL) promises privacy and distributed collaboration, its effectiveness in real-world scenarios is often hampered by the stochastic heterogeneity of clients and unpredictable system dynamics. Existing static optimization approaches fail to adapt to these fluctuations, resulting in resource underutilization and systemic bias. In this work, we propose a paradigm shift towards Agentic-FL, a framework where Language Model-based Agents (LMagents) assume autonomous orchestration roles. Unlike rigid protocols, we demonstrate how server-side agents can mitigate selection bias through contextual reasoning, while client-side agents act as local guardians, dynamically managing privacy budgets and adapting model complexity to hardware constraints. More than just resolving technical inefficiencies, this integration signals the evolution of FL towards decentralized ecosystems, where collaboration is negotiated autonomously, paving the way for future markets of incentive-based models and algorithmic justice. We discuss the reliability (hallucinations) and security challenges of this approach, outlining a roadmap for resilient multi-agent systems in federated environments.
\end{abstract}

\section{Introduction}\label{introduction}
\gls{fl} has emerged as a paradigm for training \gls{ml} and \gls{ai} models based on distributed and collaborative learning among multiple clients in a network \cite{fedavg}, where models are trained locally, and only the weights are shared with the central server to produce a global model. It enables improvements in learning by leveraging more computational resources and data at the edge, without compromising participants' privacy by sending raw data to a central server. Within this field, several challenges have been explored in recent years, such as data heterogeneity, since different sources contribute to the same global model, and computational and communication limitations, as edge devices often struggle with the intensive training of large models \cite{openproblems}.

These challenges have driven numerous research efforts aiming to address them through various strategies. For instance, client selection methods focus on the utility and necessity of each client at every round, avoiding unnecessary computational costs \cite{DESOUZA2024103462, fu2023client, 10903243, 273723}. Optimized model aggregation techniques mitigate divergence issues caused by 
heterogeneous data \cite{li2021ditto, abs-2410-22622, salehi2024flash, kang2024fednn}, while model personalization approaches tailor the global model to meet the specific needs of each client \cite{survey-pfl, 3693723, talasso2024fedsccs}. On the resource side, solutions aim to make \gls{fl} more efficient through model quantization, which reduces weight precision to 
lighten training and communication costs, model pruning, which removes redundant parameters, and additional strategies such as Parameter-Efficient Fine-Tuning (PEFT) \cite{peft, peft2} adapted to \gls{fl} for recent generative models \cite{openfedllm, fedit}.

However, despite the progress made in mitigating these challenges, the inherent dynamics, complexity, and heterogeneity of distributed \gls{fl} require holistic and adaptable methods to manage the evolving demands of the system. Approaches focusing on a single aspect, such as advanced aggregation mechanisms to improve performance, can inadvertently increase latency or computational load on client devices \cite{li2025position}. Similarly, strategies centered solely on client selection or model personalization often overlook broader impacts on energy consumption, communication costs, or total training time. Moreover, as dynamic environments evolve, client and federation requirements, such as resource constraints or data drift, may change over time, rendering static solutions ineffective \cite{CRIADO2022263}. Fixed, single-purpose approaches are hampered by the complexity and dynamism that make solving real-world problems difficult, highlighting the necessity for autonomous and flexible systems.

In this context, we present this work as a position paper to show how integrating \gls{lmagents} into \gls{fl} enables the development of flexible and adaptable solutions. \cite{yao2023react}. \gls{lmagents} are systems capable of perceiving their environment, analyzing contextual information, and making dynamic decisions, while continuously improving through feedback. We argue that this adaptive and autonomous capability enables a more holistic understanding of the \gls{fl} environment, considering aspects such as resource allocation, data heterogeneity, client performance, and dynamic participation. Additionally, we show, through a proof-of-concept, that this approach can leverage several state-of-the-art solutions at the appropriate time, based on the requirements of clients. Accordingly, the main contributions of this paper are summarized as follows:
\begin{itemize}
    \item We motivate and argue for the benefits of using \gls{lmagents} in \gls{fl}, showing how agentic systems hold great potential for addressing unresolved challenges in distributed scenarios due to their complexity and dynamicity.
    \item We discuss the key challenges and open directions for employing agentic systems in the orchestration of federated systems, which can be explored in future developments in both industry and academia.
    \item Finally, we demonstrate the viability of integrating \gls{lmagents} into \gls{fl} with a proof-of-concept\footnote{The implementation is available at \url{https://github.com/rafaeloj/k-agent}}.
\end{itemize}

The remainder of this paper is organized as follows: \Cref{sec:background} provides the context for \gls{fl} and \gls{lmagents}; \Cref{sec:agentic-fl} introduces the Agentic-FL paradigm and explains how agents can be integrated into \gls{fl}; \Cref{sec:challanges} demonstrates the main challenges of integrating these two strategies; \Cref{sec:conclusion} concludes this work.

\vspace{-0,3cm}
\section{Background}\label{sec:background}

\subsection{Federated Learning}

\gls{fl} is a distributed and collaborative paradigm for model training that enables multiple clients to contribute their computational resources and local data toward building a shared, global model \cite{fedavg}. In this setting, participants receive an initial model whose structure depends on the specific application and environment, and train it locally using their private datasets. After local training, each client sends its updated model parameters to a central server, which aggregates the client updates, typically using a weighted average proportional to the local dataset sizes, and broadcasts the updated global model back to the clients. This process is repeated over multiple communication rounds until global model convergence.

This paradigm offers several advantages, including enhanced data privacy and compliance with data protection regulations, as raw data never leaves the clients. Depending on the application, these clients can be personal smartphones, enterprises, hospitals, or financial institutions holding sensitive information. Moreover, \gls{fl} leverages distributed computational resources, parallelizing the training process and reducing dependency on a single centralized node. Its inherently distributed nature also provides access to more diverse and representative datasets, allowing better model generalization and robustness.

However, implementing \gls{fl} in real-world scenarios poses several challenges. One major issue is system heterogeneity, since edge devices often have limited computational power and communication bandwidth, constraining local training and model transmission. Another critical challenge is data heterogeneity: different clients typically collect data from distinct, non-identically distributed sources \cite{HAMEDI2025130844}. Furthermore, as a distributed computing environment, \gls{fl} must deal with dynamic participation: clients may join or leave the federation intermittently, become temporarily unavailable, or experience variations in computational resources and data distribution shifts. These factors significantly complicate the stability and convergence of federated training \cite{10.5555/3737916.3741229, 10.5555/3737916.3738181}.

Numerous approaches have been proposed to mitigate these issues, including client selection strategies, improved aggregation algorithms, personalization methods, and communication-efficient techniques \cite{273723, survey-pfl, salehi2024flash, wang2020tackling, pmlr-v119-karimireddy20a}. Nonetheless, flexible solutions that generalize across diverse operational conditions and adaptive mechanisms capable of handling environmental changes such as client availability, resource variation, and data distribution shifts remain an open research challenge \cite{HAMEDI2025130844}. In this context, this work introduces a novel approach that leverages \gls{lmagents} to enhance flexibility and adaptivity in federated learning, enabling dynamic selection and coordination of different strategies according to the current state of the federation.

\subsection{LLM-based Agents}
In recent years, \gls{lmagents} have emerged as powerful autonomous systems for adaptive decision-making and coordination across complex environments \cite{Wang2024}. Their ability to interpret context, plan actions, and communicate through natural language makes them highly suitable for distributed and dynamic systems. Within the context of \gls{fl}, agents offer a natural abstraction for both clients and the central server in tasks of coordination and orchestration, enabling analysis, reasoning, self-coordination, and adaptive responses to changing conditions such as varying client availability, resource constraints, and data shifts \cite{Wang2024}.

\gls{lmagents} typically comprise four main components \cite{10849561}. The profile defines the agent’s objectives and behaviors, constraints and limitations, and available skills, shaping its overall behavior and decision-making goals. The memory maintains contextual knowledge in both recent interactions (short-term memory) and accumulated experience (long-term memory), enabling consistent and improved decision-making over time. The planning component is responsible for decomposing goals into executable steps, reasoning about strategies, and selecting actions aligned with the current environment state. Finally, the action component executes these decisions by interacting with the environment, other agents, or external tools. Together, these elements are enhanced by LLMs, which analyze textual descriptions and environment data and generate the planning and action outputs required for agent execution.

Beyond these core components, several complementary techniques have been proposed to enhance the reasoning and operational capabilities of \gls{lmagents}. Tool usage allows agents to access external functions, APIs, or computational resources, extending their abilities beyond the language model itself \cite{10.5555/3666122.3669119}. Frameworks such as ReAct \cite{yao2023react} (Reason + Act) and \gls{cot} reasoning \cite{10.5555/3600270.3602070} enable structured problem-solving, combining step-by-step deliberation with interactive action execution. These mechanisms allow agents not only to reason abstractly but also to act in real time, query data, monitor performance, and refine their strategies based on feedback, thereby increasing autonomy and robustness in dynamic environments. Furthermore, more sophisticated solutions were developed to enhance these capabilities, such as \gls{tot}, \gls{pot}, and \gls{got} \cite{yao2023tree, ganguly2024proof, 10.1609/aaai.v38i16.29720}. Furthermore, LLM agents enable the use of reinforcement learning for online learning; this capability, coupled with the ability to solve complex problems in LLMs, makes LLM agents powerful in dynamic environments \cite{peiyuan2024agile}.

In the context of \gls{fl}, these characteristics make \gls{lmagents} particularly advantageous. Their inherent reasoning and memory mechanisms allow them to model and adapt to system heterogeneity, dynamically select or prioritize clients, personalize training strategies, and adjust aggregation methods based on the current federation state. Furthermore, the ability to use external tools enables real-time communication monitoring and performance analysis, fostering self-adaptive orchestration. Consequently, \gls{lmagents} introduce a promising paradigm for constructing flexible, context-aware, and adaptive \gls{fl} systems capable of evolving with their environments and optimizing collaboration among distributed participants.

\vspace{-0,3cm}
\section{Agent Systems for Federated Learning}\label{sec:agentic-fl}
\gls{fl} is dynamic and multivariate in nature, making its training unstable and challenging. Therefore, researchers study separately how each component of the federated system impacts training. In this section, we explain the key challenges of~\gls{fl}, considering both the server side and the client side, and how an agentic approach can be used to address these problems. We also describe how the literature tackles these challenges, explore the research opportunities that arise, and discuss how~\gls{llm} Agents can offer a promising solution.

\subsection{Server-side Agents}\label{sec:server-side}

In~\gls{fl}, systemic problems such as \textit{i) synchronization}, \textit{ii) communication overhead}, and \textit{iii) fault tolerance} make its implementation challenging. Furthermore, other factors such as \textit{iv) mitigation of bias} and \textit{v) security} are also important for maintaining a fair and secure system for users. Therefore, much research is being conducted to mitigate these impacts on~\gls{fl}.

\textbf{Synchronization.} The standard \gls{fl} protocol relies on synchronous communication, often being limited by slower clients (stragglers). While asynchronous \gls{fl} attempts to mitigate these delays via partial aggregation with time limits, it introduces the challenge of model staleness \cite{10.1109/TC.2020.2994391}. Specifically, updates from resource-limited devices may arrive late, and merging these stale weights degrades the overall model performance compared to the progress already achieved by faster clients.

In this scenario, \gls{lmagents} contribute through their adaptation to dynamic environments. Long-term memory allows contextualization of connectivity patterns: for example, by identifying that a client exhibits high latency only at specific times (e.g., at night), the agent's reasoning component can set a dynamic timeout adjusted for that round, instead of prematurely discarding it. Furthermore, global reasoning allows distinguishing isolated failures from systemic problems, such as the simultaneous unavailability of devices in the same geographic region, enabling more robust recovery strategies than simple fixed rules.

Another aspect of \gls{fl} investigates Partial Training Techniques that mitigate latency by adjusting the model size to the local hardware \cite{10.1145/3596907}, but make aggregation complex: the weighted average fails when faced with heterogeneous parameters, leaving parts of the architecture underrepresented \cite{10.1145/3596907}. In addition, the construction of submodels via ``neuron importance'' creates a systemic bias in which clients capable of training a complete architecture dominate learning, hindering the generalization capacity of the global model \cite{Wen2023}.

Agent-based methods offer a dynamic solution to this aggregation dilemma. Unlike static algorithms, a server agent can use its long-term memory to track the update frequency of each parameter or layer across runs. Upon detecting that certain sections of the model are underrepresented (due to the prevalence of weak devices), the agent's reasoning mechanism can intervene, assigning higher weights to rare updates or planning future selection to prioritize clients capable of training those neglected sections. This adaptive orchestration mitigates hardware bias and ensures the integrity of the overall model.

\textbf{Communication overhead.} Although many studies aim to reduce the communication overhead caused by \gls{fl}, this remains an open challenge \cite{10835558}. The distributed nature of the data requires multiple rounds of communication to achieve convergence during training. Consequently, sending updates over the network incurs a large payload, as these updates are directly linked to the size of the models, including large-scale architectures such as fundational models \cite{10835558}.

One way to mitigate the impact of communication overhead is through client selection. Traditionally, this is used to reduce communication bottlenecks by decreasing the number of devices sending models in each round. Essentially, client selection aims to identify the best devices within the federated system, given a specific criterion (e.g., loss, amount of data, channel size), to boost training and reduce the number of communication rounds \cite{fedavg, pmlr-v151-jee-cho22a}. However, selection algorithms that require hyperparameter optimization are vulnerable to misconfiguration problems. On the other hand, dynamic selection solutions \cite{273723, DESOUZA2024103462, 10416384}, while more efficient in federated environments, generally consider only a few selection criteria or are computationally expensive. Therefore, adaptive mechanisms that adjust to the system's needs and utilize multivariate criteria within the federated system are foundational to its evolution.

In this way, agent-based solutions are promising because they introduce a layer of meta-reasoning over client selection. By cross-referencing historical performance (long-term memory) with the current state of the network, the agent can infer which selection strategy maximizes utility at that specific moment, overcoming the rigidity of static algorithms. For example, when detecting a scenario of high hardware homogeneity but heterogeneous data, the agent can dynamically opt for Power-of-choice \cite{pmlr-v151-jee-cho22a}; while in situations where the balance between efficiency and loss is critical, it can transition to utility-based strategies such as Oort \cite{273723}. More than just selecting devices, the reasoning component allows the agent to adjust the cardinality of the selection in real time, avoiding communication bottlenecks that would go unnoticed by fixed heuristics.

In parallel, the agent can act on payload mitigation through intelligent management of compression techniques. Instead of applying uniform quantization, which can degrade the performance of capable clients, an orchestrating agent can negotiate the precision of the model individually. Using inference tools, the agent decides on sparcification or aggressive quantization only on clients with severe battery or bandwidth constraints, maintaining the integrity of the entire model for robust devices. This granularity of decision, unfeasible for traditional optimizers due to the explosion of the search space, becomes tractable through the generalization capabilities of agents.

\textbf{Fault tolerance.} In distributed systems, it is common for devices to fail, either due to connectivity issues or physical reasons such as low battery levels \cite{crawshaw2024federated, xiang2024efficient}. Thus, when selected by the server, they may be unable to perform the task or return the result, generating synchronization problems as well as delays in training. Furthermore, the server itself may fail, leading to loss of state and requiring recovery and state synchronization \cite{XU2023100595}. In this context,~\gls{lmagents} can be introduced as a complementary mechanism: when devices return online, they can send information about the reasons for their failure, which the agent can use to make informed decisions to prevent those devices from becoming underrepresented during training. At the system level, agents may also support re-establishing connections, handling error logs, and triggering recovery procedures.

\textbf{Bias.} We can divide the sources of bias into three categories ~\cite{hal-04855447, 10.1145/3735125}: \textit{i) Demographic bias:} In these cases, the imbalance of information generates an underrepresentation that is propagated to the global model if not properly addressed \cite{CRIADO2022263, ZHU2021371}. Within this context, the data used to train the models is personal; therefore, creating a single generic model for all devices may not be the best choice \cite{9743558}. \textit{ii) Systemic bias:} This is directly linked to devices with greater computational power or better connectivity, which allows them to participate in more training rounds \cite{10.1145/3735125}. Consequently, the global model tends to represent these devices more accurately than those with limited capabilities. \textit{iii) Algorithmic bias:} Many studies that aim to solve other challenges either ignore the bias they themselves introduce or explicitly rely on biased criteria. For example, some selection mechanisms use metrics such as loss to choose which device will be trained in the next round \cite{pmlr-v151-jee-cho22a, 273723}.

To address these problems, many strategies adopt an approach that produces multiple models for different groups or clients \cite{talasso2024fedsccs, survey-pfl, 9743558}. While they achieve excellent results through collaboration and personalization, challenges such as balancing generalization and personalization, avoiding negative knowledge transfer, and managing algorithmic complexity remain open \cite{Yang_2024_CVPR, NEURIPS2024_a7a6465b}. Thus, agents bring transparency and explainability to training decisions. To mitigate systemic bias, the agent can dynamically apply quantization or pruning techniques, enabling the participation of clients with limited resources without delaying federation. Furthermore, \gls{lmagents} add a qualitative and semantic analysis to combat algorithmic bias, being able to interpret client metadata (e.g., ``sensor located in a rural area'') and prioritize it to ensure demographic diversity, a nuance that simple numerical loss metrics fail to capture.

\textbf{Security.} Although \gls{fl} emerged as a privacy-oriented solution often outweighs its intrinsic security guarantees. Consequently, the focus has shifted to preventing attacks on privacy, such as data inference, and on the integrity of the model \cite{gao-etal-2025-gradient}. Classic methods like \gls{dp}, \gls{he}, and \gls{ra} mitigate these vulnerabilities, but face rigid trade-offs: \gls{dp} noise degrades accuracy, he explodes computational cost, and \gls{ra} can discard legitimate data by mistaking it for attacks \cite{10.1002/int.22818, YURDEM2024e38137}.

In this context, agents act as adaptive guardians. Unlike static policies, a security agent can dynamically manage the privacy budget ($\epsilon$), adjusting the level of injected noise based on the sensitivity of the data detected in the current round, balancing protection and utility. Furthermore, in defense against model poisoning, the \gls{lmagents}'s contextual reasoning allows it to distinguish between statistical anomalies caused by data heterogeneity (\gls{noniid} clients) and actual malicious attacks, a subtle distinction where algorithms based purely on thresholds often fail.

\subsection{Client-side Agents}

Devices within a system are responsible for model training; however, each device operates in its own context, performing additional tasks beyond model training and potentially participating in multiple federations \cite{10883172}. Significant advancements in model optimization, such as quantization and \gls{sml} enable devices to run language models \cite{belcak2025small, egashira2024exploiting}. This possibility allows for intelligent and autonomous task management on the client side. We divide client challenges into three categories: i) task manager, ii) security, and iii) training.

\textbf{Task Manager.} Edge devices execute multiple simultaneous tasks in a resource-constrained environment. While traditional schedulers work for static routines, they fail to handle the stochastic CPU and memory volatility typical of resource-constrained \gls{iot} scenarios \cite{9611880}. As the complexity of tasks increases, mechanisms capable of real-time adaptation become necessary \cite{friha2024llm}. Unlike fixed priority queues, a management agent can reason about the device's usage history to predict availability windows and orchestrate training execution, minimizing conflicts with the user experience and optimizing energy consumption.

\textbf{Security.} In classic \gls{fl} architectures, the server is assumed to be trustworthy, a premise frequently violated in real-world scenarios. While \gls{dp} mitigates risks, the application of static noise degrades the model's utility. Here, a guardian agent overcomes fixed mechanisms by dynamically managing the client's privacy budget. The agent can assess the sensitivity of the data in the current sample and inject noise proportionally, or detect gradient inversion attempts by monitoring the behavior of server requests, blocking participation if the pattern deviates from normality. This active and contextual defense is unfeasible for passive cryptographic protocols.

\textbf{Training.} Local training is the core of federation, but suffers from the double pressure of data and resource heterogeneity. Static strategies fail when attempting to optimize both simultaneously \cite{pmlr-v202-panchal23a, KANG2024110230, 9003425}. A training agent can act as a local orchestrator, capable of aligning hyperparameters not only to the architecture but also to the instantaneous state of the device. For example, upon detecting a highly skewed data distribution, the agent can apply specific regularization techniques or autonomous oversampling. Simultaneously, if resources become scarce, it unilaterally decides to apply aggressive quantization or sparcification, ensuring that the contribution is sent, even if degraded, instead of aborting the process, maximizing federation resilience.

However, while \gls{sml}s enable this local orchestration, the computational overhead and token generation costs remain critical bottlenecks for battery life and inference latency on edge devices. Therefore, developing efficient prompt strategies and managing token costs are essential research proposals for the future of \gls{lmagents}.
\vspace{-0,3cm}
\section{Challenge and Future Directions}\label{sec:challanges}
\begin{figure*}[h]
    \centering
    \includegraphics[width=\textwidth]{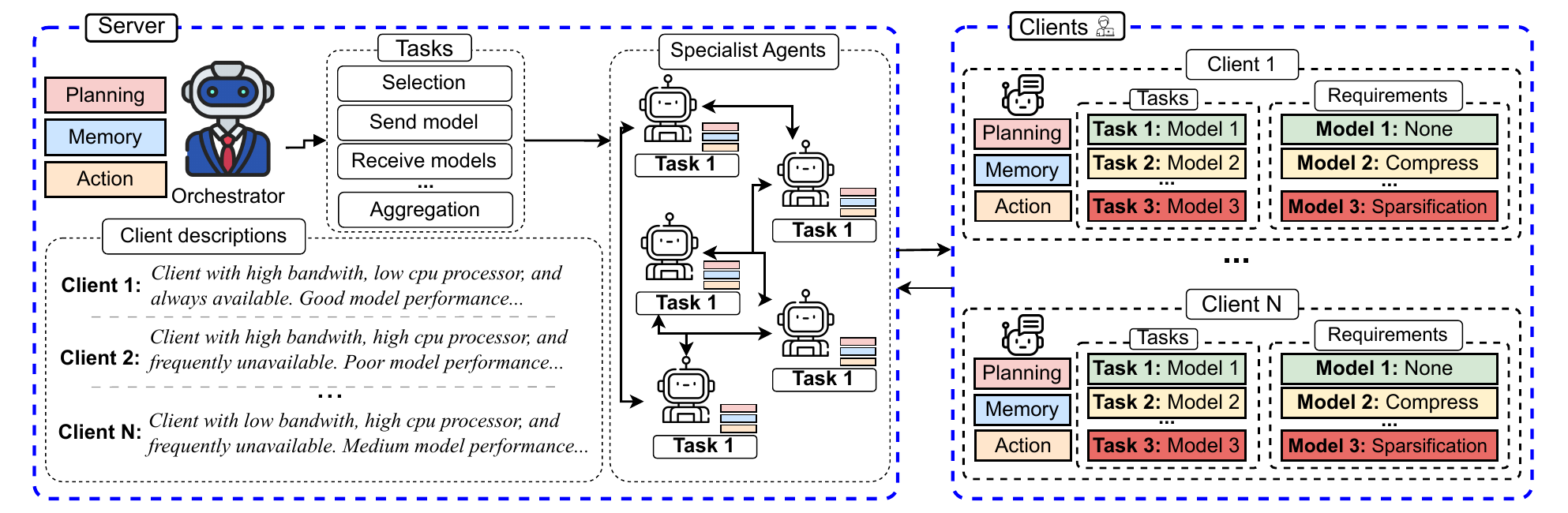}
    \caption{AgenticFL Paradigm: shows how an agent or a multi-agent system can be introduced into the~\gls{fl} pipeline.}
    \label{fig:agent-pipeline}
\end{figure*}

\gls{lmagents} have proven to be a promising methodology for mitigating~\gls{fl} problems, but their introduction also raises issues that require further study. Therefore, in this subsection, we discuss the challenges that emerge in this area as well as the research opportunities that follow.

\Cref{fig:agent-pipeline} illustrates how an agentic architecture can be built within a distributed environment. Inside the server, a multi-agent system is responsible for managing the distributed environment, while on the client side, a \gls{lmagents} manages local tasks. The server-side agent acts based on client states, activity histories, and previous decisions, determining which tasks should be executed next and identifying their dependencies. These tasks are then forwarded to specialist agents responsible for solving each task created by the orchestrator, and these specialists may also exchange information with one another. This approach helps to modularize and isolate each stage of the distributed pipeline. Thus, we are not limited to selection and aggregation but can also customize models to meet device-specific needs. For example, if the system detects clients with limited computational resources, the orchestrator agent can update their descriptions upon observing slow responses and create a quantization task before sending the model. Similarly, if a client is underperforming, the system can request additional training time. These agents can even maintain their own communication channels with the server, informing it of the requirements associated with each task or model. In this way, the client can define its level of contribution and the conditions under which it will train the model in coordination with the orchestrating agent. However, there are a number of problems in linking these two areas, such as:

\textbf{Reliability and Hallucination}. Despite their sophistication, \gls{lmagents} are still susceptible to hallucinations \cite{yang2025minimizing}. In the context of \gls{fl}, these flaws can lead the agent to misinterpret the utility of a client, resulting in the repeated selection of ineffective devices or the exclusion of significant contributors. From the client's perspective, agents orchestrating local training may hallucinate inappropriate hyperparameters; when aggregated, these updates can cause unintentional model poisoning, degrading the global model and hindering federation convergence. Therefore, the use of safety mechanisms, such as guardrails or structured outputs, to validate values and impose strict operational constraints becomes imperative.

\textbf{Agent Components.} Agent systems can be constructed in several ways, and their architectures are directly shaped by the problems they aim to solve. This applies equally to~\gls{fl}; thus, it is necessary to investigate how each agent component can contribute to the federated learning process, from both the server and client perspectives.

Within~\gls{fl}, actions performed in the past can be very useful for understanding the current state of the federation and for determining what needs to be done to achieve the expected result. Thus, the memory component is fundamental for building an agent in this context. However, there are different ways to represent information to the agent, such as using text, embeddings stored in databases, or structured lists. Finding the best representation for each step and each piece of information within the distributed system is challenging, as is defining how to display client metadata to the agent, namely, how to efficiently leverage the context windows of language models without sending unnecessary information. Furthermore, the large amount of information generated must be stored in a way that enables efficient and practical retrieval, making the study of storage methods and retrieval mechanisms essential.

Within a distributed system, the order and manner in which tasks are performed are important for conserving resources and efficiently utilizing system components. Therefore, planning task sequences based on both past actions and the current state, while aiming for short- and long-term gains, is also challenging. Understanding how to enable the agent to plan its actions using~\gls{cot},~\gls{tot}, and similar techniques to ensure reliability and robustness in its decisions will make its behavior more effective. Furthermore, the planning and reasoning process generally generates a large number of tokens, linked to the information produced by the federated system; therefore, it is important to identify mechanisms to mitigate this impact.

\textbf{Security.} Beyond the problems inherent to~\gls{fl},~\gls{lmagents} introduce additional vulnerabilities that require investigation. Miao Yu et al.\ summarize these issues as follows: the~\gls{lm} itself is vulnerable to \textbf{jailbreaking}, that is, methods used to bypass security mechanisms; \textbf{prompt injection}, which attempts to alter the agent's behavior by injecting malicious prompts into the~\gls{llm}; and \textbf{backdoors}, which insert triggers during training to produce attacker-specified outputs. In addition, it is necessary to prevent clients from poisoning the model's memory with false information through \textbf{memory poisoning}, or from obtaining improper information that leads to \textbf{privacy leakage}~\cite{10.1145/3711896.3736561}. Techniques capable of handling attempts at manipulation or abuse of tools used by the agent are fundamental to preventing overload and serious impacts on the federated system. In scenarios where many clients deploy agents, risks such as cooperative agent attacks or infection attacks that aim to corrupt agent behavior naturally arise, and existing studies on collaborative or topological defense strategies may be leveraged to address these threats.

\textbf{Federated pipeline.} Today, there are different~\gls{fl} paradigms, such as asynchronous, hierarchical, horizontal, and vertical federation~\cite{10.1561/2200000083}. With the advancement of agents, we can envision a framework in which agents distributed across clients can request federation when necessary. In this way, a client-side agent could, through internal mechanisms, determine when its model is no longer adequate or is becoming outdated and subsequently request a new model from the server, which would initiate a federation round after receiving a sufficient number of such requests. This new pipeline introduces adaptability to the models and flexibility to the training process, enabling federation to occur only among clients that require it.

\textbf{Exploring new metrics with Code Agents.} Traditional metrics such as accuracy and loss are often insufficient for comprehensive decision-making, and assessing client utility more holistically remains a challenge. Code Agents can propose and implement solutions that evaluate new metrics and combine information to construct previously undefined ones. For example, they can identify a group of disadvantaged clients and generate a minority fairness metric, enabling the system to optimize its actions to benefit this group in subsequent rounds.

\textbf{Scalability and Contextual Limits} Scaling the federation to thousands of clients inevitably hits the physical limits of \gls{lm} context windows, demanding architectural innovations to reduce information overload. To address this, we envision three complementary directions. First, \gls{rag} can dynamically fetch only relevant client metadata, creating similarity clusters rather than processing the entire network at once. Second, hierarchical multi-agent architectures can divide the workload, where local sub-agents analyze specific client subgroups and forward only condensed, high-value candidates to a central orchestrator. Finally, integrating deterministic filtering tools can pre-categorize clients based on heuristics, allowing the agent to focus its reasoning solely on a pre-selected, manageable pool of candidates.

\textbf{Strategic Modeling and Mechanism Design:} In a decentralized ecosystem, clients are naturally driven by local constraints and objectives, which may lead them to underreport their available resources to save battery, or overstate their data quality to bias the global model toward their specific needs. To design robust defenses against selfish or malicious clients, Agentic-FL can be formalized through Mechanism Design. Each client $i$ has a truly private type, denoted by $\theta_i = (b_i, c_i, q_i, bw_i)$, where $b_i$ is the battery level, $c_i$ the computational cost, $q_i$ the data quality, and $bw_i$ the bandwidth. The client reports a type $\hat{\theta}_i$ to the Orchestrator Agent, and can report incorrect values ($\hat{\theta}_i \neq \theta_i$) to gain advantages. The server acts as the mechanism designer, defining an allocation rule $w_i(\hat{\theta}_i)$ (the weight or probability of selection in the round) and a penalty or reputation function $\Pi$. The utility of the strategic client is given by:
\begin{equation}
    U_i(\hat{\theta}_i | \theta_i) = v_i \cdot w_i(\hat{\theta}_i) - C(\theta_i) - \Pi(\text{audit}(\hat{\theta}_i))
\end{equation}

where $v_i$ is the value the client assigns to influence in the global model and $C(\theta_i)$ is its actual training cost. While classical Mechanism Design relies on rigid numerical inputs to solve this equation, Agentic-FL introduces \gls{lmagents} to operationalize these dynamics in unstructured environments. Here, the Orchestrating Agent uses contextual reasoning to dynamically infer the reported types $\hat{\theta}_i$ from complex metadata and behavioral logs. The open challenge for the community is the transition from static mathematical formulations to \gls{lmagents} driven mechanisms, where agents autonomously negotiate incentives, audit contributions, and adjust the value of $\Pi$ in real time based on semantic analysis, rather than just numerical limits.

The integration of \gls{lmagents} into \gls{fl} transcends mitigating isolated challenges; it signals a paradigm shift from centralized and static orchestration to autonomous federated ecosystems. While traditional \gls{fl} treats clients as passive ``workers'' our long-term vision projects a scenario where agents act as economic brokers in a decentralized knowledge marketplace. In this future, the training pipeline is no longer triggered by a central server but negotiated peer-to-peer: clients initiate federations based on local needs (e.g., ``I need to update my night vision model'') and negotiate participation based on new metrics of utility and fairness, not just accuracy. This evolution transforms FL from a distributed optimization process into a complex mechanism for incentive negotiation and on-demand collaboration, opening frontiers for the intersection between Game Theory and Machine Learning.

\vspace{-0,3cm}
\section{Proof-of-Concept}\label{sec:poc}
This section illustrates the potential of use llm agents to define the number of K clients to select. First, we describe the setup and methods to construct the K-Agent. Second ,we evaluate the results of experiments.

\textbf{Setup.}
The simulation of \gls{fl} was implemented using the Flower framework, utilizing the CIFAR-10 and MNIST datasets under a Dirichlet distribution ($\alpha = 0.1$) to create a severely Non-IID scenario with 25 clients and 50 rounds of communication. The proposed method, K-Agent, was developed via LangGraph and Ollama, testing three models: Qwen3 8b, Llama3.2 3b, and Llama3.1 8b. This involved training a CNN optimized by SGD and comparing it to established baselines such as Random Selection, Round Robin, Pow-of-Choice, and Oort. Effectiveness was validated by averaging three runs based on two fundamental metrics: Distributed Accuracy, to measure overall performance; Selection Time (ST), which adds the LLM inference to the selection algorithm, which accounts for the network overhead accumulated by client selection using three techniques: \gls{cot}, Few-Shot, and Description Only (DO).

\textbf{Results.} \Cref{tab:tempo-de-selecao} table presents the accuracy, selection time (ST), and average $K$ resulting from different combinations of models and prompts for MNIST and CIFAR10. It can be observed that the \textit{Qwen3 8b} model tended to choose lower $K$ values, but with a higher standard deviation. This variability allowed for higher accuracies, but with higher ST. While for MNIST we can see that \textit{Llama3.2 3b} achieved results similar to the other models with a lower ST due to shorter and more objective thought chains. However, despite this, in some cases the agent gets lost in the reasoning, requiring other calls and increasing its time, as demonstrated in the Oort using \gls{cot}.\footnote{\gls{cot} benefits models large enough to generalize reasoning lines \textit{step-by-step} as evidenced by Wei et al. \cite{wei2022chain}.}

Regarding prompt techniques, \gls{cot} showed a predominantly lower TS. Some exceptions occurred with \textit{Llama3.1 8b}, where a simple description was faster; however, this configuration resulted in static $K$ values for PoC and Random, reducing the agent's adaptive effectiveness.

\begin{table}[]
\centering
\caption{Impact of LLM and Prompt on Selection and Performance}
\label{tab:tempo-de-selecao}
\renewcommand{\arraystretch}{1.2}
\resizebox{\textwidth}{!}{%
\begin{tabular}{m{1.5cm}cccccccccc}
\toprule
\multicolumn{11}{c}{\textbf{Cifar 10}}                                                                                                                                                                                                                            \\ \bottomrule
\multirow{2}{*}{\textbf{Models}}     & \multirow{2}{*}{\textbf{Prompt}} & \multicolumn{3}{c}{\textbf{PoC}}                         & \multicolumn{3}{c}{\textbf{Random}}                      & \multicolumn{3}{c}{\textbf{Oort}}                        \\ \cmidrule{3-11} 
                                      &                                          & \textbf{K avg} & \textbf{Acc.} (\%) & \textbf{ST} (s)       & \textbf{K avg} & \textbf{Acc.} (\%) & \textbf{ST} (s)       & \textbf{K avg} & \textbf{Acc.} (\%) & \textbf{ST} (s)       \\ \midrule
\multirow{3}{*}{\textbf{Qwen3}}    & \textbf{DO}                     & 7$\pm$5             & 38$\pm$13            & 3,79           & 6$\pm$5             & \textbf{39$\pm$11}   & 4,71          & 12$\pm$6            & \textbf{39$\pm$14}   & 4,45          \\
                                      & \textbf{Few-Shot}                        & 6$\pm$4             & 38 $\pm$13           & 1,43          & 5$\pm$4             & 38$\pm$15            & 2,28          & 9$\pm$5             & 38$\pm$15            & 2,96          \\
                                      & \textbf{CoT}                             & 7$\pm$5             & \textbf{39$\pm$13}   & 1,79          & 5$\pm$4             & 37$\pm$14            & 1,65          & 9 $\pm$4            & 39 $\pm$13           & 2,53          \\
\multirow{3}{*}{\textbf{Llama3.1}} & \textbf{DO}                     & 10$\pm$0            & 37$\pm$13            & \textbf{0,18} & 10$\pm$0            & 38$\pm$10            & 0,38          & 9$\pm$2             & 38$\pm$13            & \textbf{0,19} \\
                                      & \textbf{Few-Shot}                        & 8$\pm$6             & 37$\pm$13            & 3,21          & 10$\pm$7            & 38$\pm$12            & 1,88          & 8$\pm$4             & 38+15             & 3,92          \\
                                      & \textbf{CoT}                             & 8$\pm$2             & 38 $\pm$12           & 0,53          & 8$\pm$2             & 37$\pm$14            & 0,51          & 8 $\pm$1            & 39 $\pm$12           & 0,57          \\
\multirow{3}{*}{\textbf{Llama3.2}} & \textbf{DO}                     & 10$\pm$0            & 36$\pm$13            & 0,29          & 9$\pm$1             & 36$\pm$14            & 0,21 & 9$\pm$1             & 39$\pm$11            & 0,31          \\
                                      & \textbf{Few-Shot}                        & 5$\pm$0             & 35$\pm$13            & 0,31          & 10$\pm$7            & 36$\pm$11            & 1,88          & 9$\pm$1             & 38$\pm$14            & 0,27          \\
                                      & \textbf{CoT}                             & 9$\pm$1             & 35+15             & 0,20          & 8$\pm$2             & 36$\pm$13            & \textbf{0,15}          & 8$\pm$1             & 38$\pm$13            & 0,22          \\ \bottomrule
\multicolumn{11}{c}{\textbf{MNIST}}                                                                                                                                                                                                                               \\ \bottomrule
\multirow{2}{*}{\textbf{Models}}     & \multirow{2}{*}{\textbf{Prompt}} & \multicolumn{3}{c}{\textbf{PoC}}                         & \multicolumn{3}{c}{\textbf{Random}}                      & \multicolumn{3}{c}{\textbf{Oort}}                        \\ \cmidrule{3-11} 
                                      &                                          & \textbf{K avg} & \textbf{Acc.} (\%) & \textbf{ST} (s)       & \textbf{K avg} & \textbf{Acc.} (\%) & \textbf{ST} (s)       & \textbf{K avg} & \textbf{Acc.} (\%) & \textbf{ST} (s)       \\ \midrule
\multirow{3}{*}{\textbf{Qwen3}}    & \textbf{DO}                     & 7$\pm$6             & 95,97$\pm$3          & 4,00          & 6$\pm$5             & \textbf{96,90$\pm$1} & 8,65          & 11$\pm$6            & \textbf{97,06$\pm$1} & 4,45          \\
                                      & \textbf{Few-Shot}                        & 7$\pm$5             & 96,68$\pm$2          & 1,67           & 5$\pm$4            & \textbf{96,32$\pm$2} & 1,74          & 8$\pm$6             & \textbf{97,22$\pm$1} & 1,97          \\
                                      & \textbf{CoT}                             & 7$\pm$5             & \textbf{96,50$\pm$2} & 1,63          & 8$\pm$2             & 95,25$\pm$2          & 0,18          & 9$\pm$6             & \textbf{97,05$\pm$1} & 2,21          \\
\multirow{3}{*}{\textbf{Llama3.1}} & \textbf{DO}                     & 10$\pm$0            & \textbf{96,04$\pm$3} & \textbf{0,17} & 10$\pm$2            & \textbf{96,66$\pm$2} & 0,18          & 9$\pm$2             & \textbf{97,05$\pm$1} & 0,19          \\
                                      & \textbf{Few-Shot}                        & 10$\pm$1            & 95,76$\pm$2          & 0,16         & 6$\pm$1              & 95,86$\pm$3          & 0,34          & 9$\pm$1             & \textbf{97,21$\pm$1} & 0,33          \\
                                      & \textbf{CoT}                             & 8$\pm$2             & \textbf{96,64$\pm$2} & 0,26          & 8$\pm$2             & 95,89$\pm$2          & 0,21          & 5$\pm$1             & \textbf{97,09$\pm$1} & 0,28          \\
\multirow{3}{*}{\textbf{Llama3.2}} & \textbf{DO}                     & 10$\pm$0            & 95$\pm$3             & 0,31          & 9$\pm$1             & \textbf{96,64$\pm$1} & 0,20          & 9$\pm$1             & \textbf{97,22$\pm$1} & 0,29          \\
                                      & \textbf{Few-Shot}                        & 9$\pm$2             & \textbf{96,26$\pm$2} & \textbf{0,16} & 7$\pm$2             & \textbf{96,13$\pm$2} & \textbf{0,14} & 8$\pm$1             & \textbf{97,00$\pm$1} & \textbf{0,16}  \\
                                      & \textbf{CoT}                             & 9$\pm$1             & \textbf{96,13$\pm$2} & 0,19          & 8$\pm$2             & \textbf{96,25$\pm$2} & 0,18         & 5$\pm$1             & \textbf{96,77$\pm$1} & 2,21          \\ \bottomrule
\end{tabular}%
}
\end{table}

Combined with sophisticated techniques such as~\gls{tot}, long-term and short-term memory, and even advanced retrieval techniques and tools, agents represent promising approaches for the dynamic management of federated systems.

\vspace{-0,3cm}
\section{Conclusion}\label{sec:conclusion}
In this work, we introduce Agentic-FL, a paradigm shift that evolves \gls{fl} from a static optimization process to an autonomous trading ecosystem. We argue that the inherent complexity of distributed scenarios, characterized by non-IID data heterogeneity, resource volatility, and security risks, demands more than fixed algorithms; it demands cognitive adaptation.

Our analysis demonstrated that the integration of LLM-based Agents offers a robust solution on two fronts: on the Server, agents act as orchestrators capable of combating systemic bias through contextual aggregation and dynamic selection; on the Client, agents operate as ``local gatekeepers'' managing privacy budgets and adapting training in real time to ensure the inclusion of resource-limited devices.

By mitigating barriers to entry for heterogeneous clients and increasing resilience against attacks, Agentic-FL not only improves training efficiency but also promotes a fairer and more representative system. Looking to the future, we envision these agents as the forerunners of decentralized market models, where collaboration ceases to be merely a computational protocol and becomes an intelligent, self-managed economic exchange.


\subsubsection*{Acknowledgments}
This project was supported by the brazilian Ministry of Science, Technology and Innovations, with resources from Law nº 8,248, of October 23, 1991, within the scope of PPI-SOFTEX, coordinated by Softex and published Arquitetura Cognitiva (Phase 3), DOU 01245.003479/2024-10 and was partially sponsored by CNPq grant 407192/2025-5.


\bibliography{iclr2026_conference}
\bibliographystyle{iclr2026_conference}

\appendix
\section{Appendix}
\subsection{Semantic analysis of K-Agent results}

\begin{figure}[!ht]
    \centering
    \includegraphics[width=\linewidth]{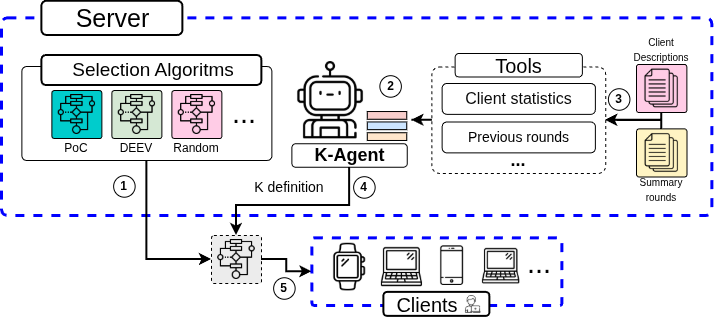}
    \caption{Pipeline of K-Agent}
    \label{fig:k-agent-pipeline}
\end{figure}

\Cref{fig:k-agent-pipeline} depicts the overview. K-Agent is implemented on the central server and is triggered at the beginning of each communication round $t$, consisting of three interconnected modules: Plan, Memory, and Action, as illustrated in Figure 1. The first step of the process (1) consists of defining the selection algorithm that will be used during training. This definition is performed before the start of training, and the same algorithm is used in all rounds. In the next step (2), in each communication round, the agent is triggered to define the ideal value for $K$, which operates on the client and federation metadata highlighted in step (3). In this part, the tools allow the agent to collect information about the training metadata of each client (e.g., training time, model performance, link quality, etc.) and the federation as a global average of each attribute. Then, in step (4), the agent inserts the hyperparameter into the defined algorithm, and the system executes the solution. The server then, in step (5), sends the global model to the selected clients, continuing the traditional flow.

\begin{figure}[ht]
    \centering
    \includegraphics[width=\linewidth]{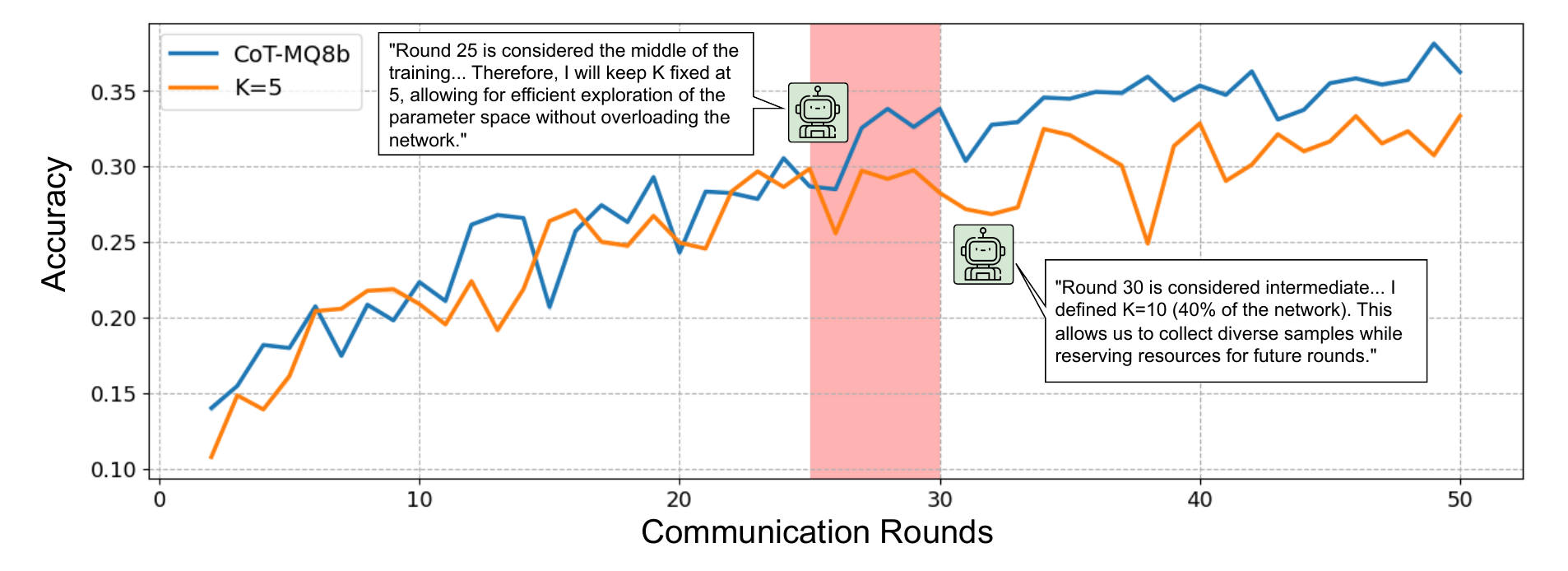}
    \caption{Convergence of \gls{poc}: The interval highlights the dynamic adaptation of the agent's strategy.}
    \label{fig:convergencia}
\end{figure}

\Cref{fig:convergencia} illustrates that the superiority of K-Agent transcends metric performance, being based on its contextual adaptability. By diverging from the fixed strategy in round 25, the agent demonstrates a dynamic perception of the federation's state that static methods lack. When the system detects imminent instability, maintaining the value of $K=5$ acts as a variance control mechanism, stabilizing the global gradient, before resuming a more aggressive exploration of clients after round 30. This smooth transition between conservatism and exploration is what ensures not only faster convergence but also superior operational robustness in heterogeneous environments.

Additionally, the agent-based architecture introduces a layer of auditability and explainability. Through the analysis of reasoning logs, it is possible to decode the causal logic behind each change in the value of $K$, transparently. This traceability allows developers to audit the model's heuristics in real time, validating whether the choice for greater customer diversity in advanced rounds was motivated by model stability or by specific performance metrics. Thus, K-Agent establishes a new paradigm where training effectiveness is inseparable from clarity about the system's behavior.

\subsection{Compare single llm vs agent}\label{sec:poc}
To illustrate some of the use cases discussed in the previous sections, we designed some experiments with the agentic system. We decided to conduct an experiment comparing the agentic solution with~\gls{fedavg}. We compared three different techniques: i) using only~\gls{llm} with all client descriptions inserted in the model's context window; ii) an agent with ReAct architecture ~\cite{yao2023react} with five tools. These two solutions were compared with the classic~\gls{fedavg} solution in order to understand the potential that agents can achieve.

\textbf{Setup:} The models used were gpt-4o-mini from the OpenAI API. For the experiments, since the objective of this experiment is simply to understand the potential of~\gls{llm}s, we used a simple MNIST database with a DNN.

\textbf{Experiment 1:} To understand the potential of language models for client selection, we conducted an experiment comparing three strategies: i)~\gls{fedavg}, i.e., random selection; ii) Selection using an~\gls{llm} as a selector, for which all client information and the sequence of instructions for selection are input. iii) In addition, we used a simple ReAct architecture that queries client information through three tools (\textit{filter}, \textit{get\_stats}, \textit{get\_info\_by\_cid}). The simulation was performed with 10 clients for 25 rounds in a~\gls{noniid} scenario. During the experiments, we tested between 5 and 50 clients with dynamic and fixed selection, which will be explained during each experiment.

\begin{figure}[!h]
    \centering
    \includegraphics[width=0.5\linewidth]{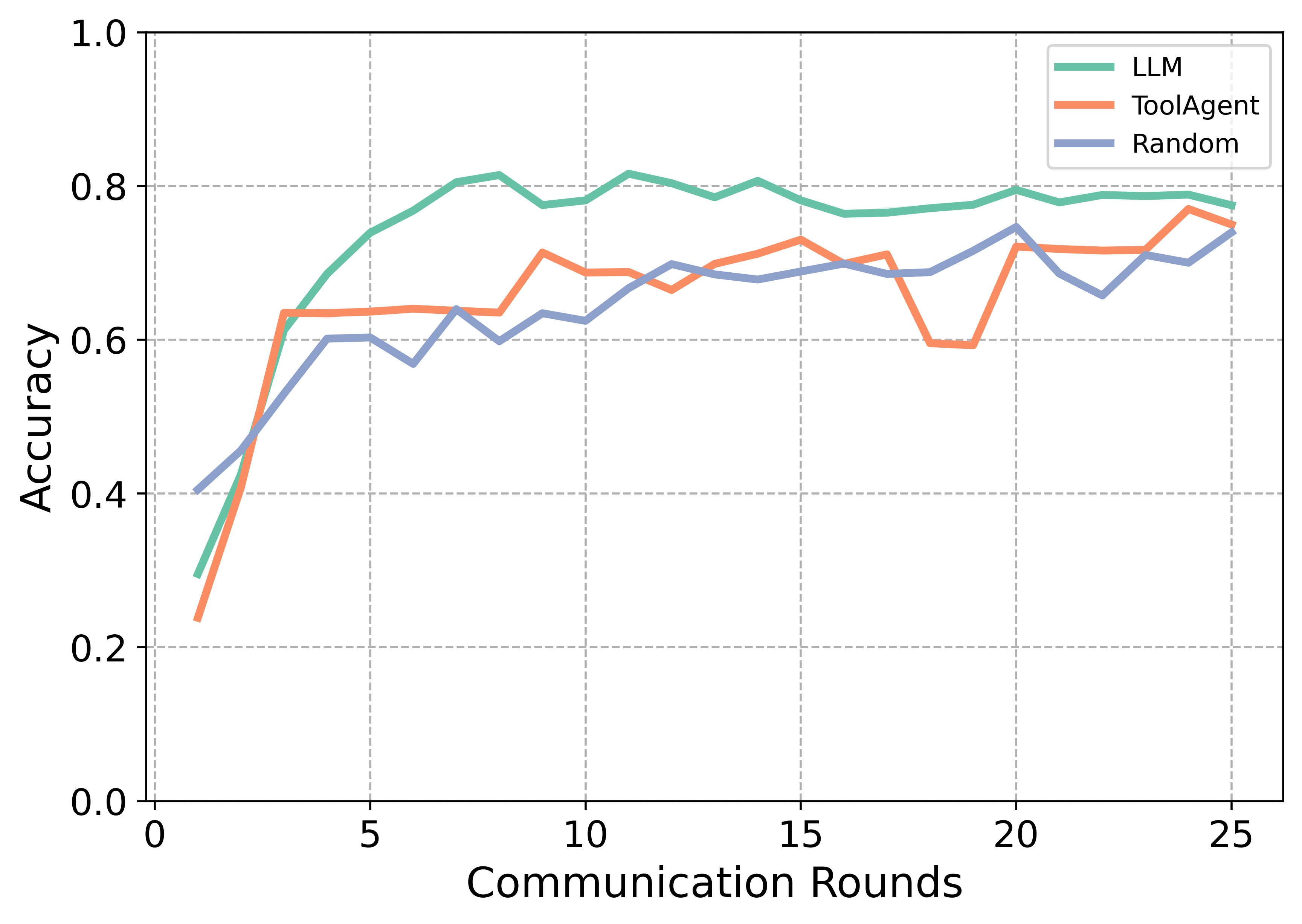}
    \caption{Experiment comparing raw~\gls{llm} to random selection and ToolAgent.}
    \label{fig:raw_llm}
\end{figure}

As we can see in~\Cref{fig:raw_llm}, the best-performing method was~\gls{llm}. This is based on the model having the full context of client information, while the agent must decide how to call the functions to filter clients and then perform the analysis of which ones to select. By analyzing the justifications for both strategies, we realized that they were able to distinguish between training and evaluation performance, in addition to using it to their advantage, selecting clients with poorer performance in the evaluation.~\gls{llm} predominantly took into account other attributes, such as training time, which the Tool agent did not consider in any round, despite the information being provided. The Tool agent had confusion about the meaning of some attributes. Both considered criteria such as diversity and fairness, but the Tool agent demonstrated greater coherence in its choices, as clients with lower performance were selected to maintain diversity in training.

\renewcommand{\arraystretch}{1.5}
\begin{table}[!ht]
\caption{The number of tokens used according to the increase in the number of clients by~\gls{llm}.}
\resizebox{\linewidth}{!}{
    \begin{tabular}{ccccccc}
    \hline
    \textbf{N° Clients}          & \textbf{Approachs} & \textbf{Completion Tokens} & \textbf{Prompt Tokens}   & \textbf{Total Cost (\$)}              & \textbf{Total Tokens}    & \textbf{Accuracy}                \\ \hline
    \multirow{2}{*}{\textbf{5}}  & LLM                & 178 $\pm$ 52               & \textbf{1336 $\pm$ 15}   & \textbf{0.000308 $\pm$ 0.000031} & \textbf{1515 $\pm$ 54}   & 0.766264 $\pm$ 0.105471          \\
                                 & Tool Agent         & \textbf{164 $\pm$ 122}     & 6193 $\pm$ 1963          & 0.000804 $\pm$ 0.000241          & 6358 $\pm$ 2072          & \textbf{0.792612 $\pm$ 0.101721} \\ \hline
    \multirow{2}{*}{\textbf{10}} & LLM                & \textbf{218 $\pm$ 56}      & \textbf{2259 $\pm$ 28}   & \textbf{0.000470 $\pm$ 0.000035} & \textbf{2478 $\pm$ 64}   & 0.693686 $\pm$ 0.066431          \\
                                 & Tool Agent         & 289 $\pm$ 218              & 8367 $\pm$ 3103          & 0.001149 $\pm$ 0.000431          & 8657 $\pm$ 3297          & \textbf{0.785554 $\pm$ 0.085655} \\ \hline
    \multirow{2}{*}{\textbf{25}} & LLM                & 254 $\pm$ 67               & \textbf{5035 $\pm$ 73}   & \textbf{0.000908 $\pm$ 0.000041} & \textbf{5289 $\pm$ 97}   & 0.637390 $\pm$ 0.090126          \\
                                 & Tool Agent         & \textbf{248 $\pm$ 244}     & 8056 $\pm$ 3275          & 0.001080 $\pm$ 0.000484          & 8305 $\pm$ 3493          & \textbf{0.882115 $\pm$ 0.146230} \\ \hline
    \multirow{2}{*}{\textbf{50}} & LLM                & 241 $\pm$ 68               & 9652 $\pm$ 145              & 0.001593 $\pm$ 0.000044          & 9894 $\pm$ 154           & 0.728411 $\pm$ 0.126316          \\
                                 & Tool Agent         & \textbf{219 $\pm$ 223}     & \textbf{7185 $\pm$ 2343} & \textbf{0.000961 $\pm$ 0.000325} & \textbf{7404 $\pm$ 2539} & \textbf{0.797055 $\pm$ 0.138048} \\ \hline
    \end{tabular}
}
\label{tab:n-clients}
\end{table}

However, using language models alone is not scalable, as demonstrated in the~\Cref{tab:n-clients}. For this experiment, we ran each approach three times, increasing the number of clients, to understand the scalability of the strategies. As we can see, as we increase the number of clients, the number of tokens increases, along with its cost for the pure method~\gls{llm}. While the agent tool starts with a high value, the growth is slower. With 50 clients, the cost of~\gls{llm} exceeds that of the agent. Although the standard deviation indicates a similar cost, the potential for improvement is promising with the improvement of the action and query tools, providing information intelligently and reducing costs. In addition to the cost reduction, we also observed a performance gain when using the models coupled to tools when we scale the number of clients, as the model becomes less confused.

Combined with sophisticated techniques such as~\gls{tot}, long-term and short-term memory, and even advanced retrieval techniques and tools, agents represent promising approaches for the dynamic management of federated systems.

\end{document}